\documentclass[fleqn,usenatbib]{mnras}

\usepackage{newtxtext,newtxmath}
\usepackage[T1]{fontenc}

\DeclareRobustCommand{\VAN}[3]{#2}
\let\VANthebibliography\thebibliography
\def\thebibliography{\DeclareRobustCommand{\VAN}[3]{##3}\VANthebibliography}

\usepackage{graphicx}	% Including figure files
\usepackage{amsmath}	% Advanced maths commands

\usepackage{amssymb}	% Extra maths symbols
\newcommand{\angstrom}{\text{\normalfont\AA}}
\usepackage{multirow}
\usepackage{array}
\newcolumntype{L}[1]{>{\raggedright\let\newline\\\arraybackslash\hspace{0pt}}m{#1}}
\newcolumntype{C}[1]{>{\centering\let\newline\\\arraybackslash\hspace{0pt}}m{#1}}

\newcommand{\erg}{erg cm$^{-2}$ s$^{-1}$}
\newcommand{\lum}{erg s$^{-1}$}
\newcommand{\lmc}{LMC X--4 }
\newcommand{\astro}{\emph{AstroSat }}

\title[AstroSat observation of LMC X--4]{Broadband mHz QPOs and spectral study of \lmc with \astro}

\author[R. Sharma et al.]{
Rahul Sharma,$^{1}$\thanks{E-mail: rsharma@rri.res.in (RS)}
Chetana Jain,$^{2}$\thanks{E-mail: chetanajain11@gmail.com (CJ)}
Ketan Rikame,$^{1,3}$
and Biswajit Paul$^{1}$
\\
$^{1}$Raman Research Institute, C.V. Raman Avenue, Sadashivanagar, Bengaluru 560080, Karnataka, India \\
$^{2}$Hansraj College, University of Delhi, Delhi 110007, India\\
$^{3}$Department of Physics and Electronics, CHRIST (Deemed to be University), Bengaluru 560029, Karnataka, India\\
}

\date{Accepted XXX. Received YYY; in original form ZZZ}

\pubyear{2022}

\begin{document}

\label{firstpage}
\pagerange{\pageref{firstpage}--\pageref{lastpage}}
\maketitle

\begin{abstract}

We report the results of broadband timing and spectral analysis of data from an \astro observation of the High Mass X-ray binary LMC X--4. The Large Area X-ray Proportional Counter (LAXPC) and  Soft X-ray Telescope (SXT) instruments on-board the \astro observed the source in August 2016. A complete X-ray eclipse was detected with the LAXPC. The 3--40 keV power density spectrum showed the presence of coherent pulsations along with a $\sim$26 mHz quasi-periodic oscillation feature. 
The spectral properties of \lmc were derived from a joint analysis of the SXT and LAXPC spectral data. The  0.5--25 keV persistent spectrum comprised of an absorbed high energy cutoff power law with photon index of $\Gamma \sim$ 0.8 and cutoff at $\sim$16 keV, a soft thermal component with kT$_{BB} \sim$ 0.14 keV and Gaussian components corresponding to Fe K$_\alpha$, Ne \textsc{ix} and Ne \textsc{x} emission lines. Assuming a source distance of 50 kpc, we determined 0.5--25 keV luminosity to be $\sim 2 \times  10^{38}$ erg s$^{-1}$.

\end{abstract}


%%%%%%%%%%%%%%%%%%%%%%%%%%%%%%%%%%%%%%%%%%%%%%%%%%

%%%%%%%%%%%%%%%%% BODY OF PAPER %%%%%%%%%%%%%%%%%%

\section{Introduction}

\lmc is a highly luminous, eclipsing high mass X-ray binary (HMXB) which is located in the Large Magellanic Cloud \citep{Giacconi72}. \lmc exhibits variations in X-ray flux over a wide range of time scales. It harbours a 1.25 M$_\odot$ neutron star spinning at a period of 13.5 s and accreting matter from a 14th magnitude OB star \citep{Pakull76, Pesch76, Kelley83, Meer07}. The binary has an inclination angle of 59$^{\circ}$ \citep{Inoue19}. The pulsar orbits its donor in an almost circular orbit. The X-ray eclipse lasts for a duration of about 5 hr and recurs every $\sim$1.4 d \citep{Li78, White78}. The orbital period is known to decay at a rate $\left(\dot{P}_{\rm orb}/P_{\rm orb}\right)$ of about 10$^{-6}$ yr$^{-1}$ \citep{Levine00, Naik04, Falanga15}. Over and above this, a 30.5 d intensity variation is also seen in this system \citep{Lang81, Paul2002, Molkov2015}. It is attributed to a precessing tilted accretion disc which periodically obscures direct X-ray emission from the compact object. Flaring episodes (of duration up to $\sim$ 45 min) have been reported in LMC X--4 \citep{Epstein77, Levine91, Beri2017}. These large X-ray flares show  quasi-periodic variability at frequencies of $\sim 0.65-1.35$ and $\sim 2-20$ mHz \citep{Moon2001}. Recently, $\sim 27$ mHz quasi-periodic oscillation (QPO) was detected with \emph{XMM}-Newton during the persistent emission \citep{Rikame22}.

The spectral variations in \lmc are equally interesting. The 0.1-100 keV spectrum of \lmc is described by a hard power law with a high-energy cutoff, a strong iron emission line and a soft X-ray excess originating due to reprocessing of hard X-rays by the inner accretion disc \citep{Levine91, Woo96, Barbera01, Paul02, Naik03, Hickox04, Shtykovsky17}. The spectral properties are known to significantly depend on the high/low super-orbital phase \citep{Naik03}. There exists a positive correlation between the 7--25 keV flux and the flux of the iron emission line. The spectrum is observed to be relatively flat during the low state in comparison to the high state \citep{Naik03}. 

In this paper, we present the results obtained from an \astro - LAXPC and SXT observation of \lmc made during 2016 August. The paper is organised in the following way. Section §\ref{sec:obs} gives a description of the observation and the data reduction procedure. In Section §\ref{sec:result}, the results from timing and spectral analysis of \lmc have been presented. Our findings have been discussed in Section §\ref{sec:discuss}.

\section{Observations}
\label{sec:obs}

\begin{table*}
\caption{Log of X-ray observations.}
\centering
\resizebox{2\columnwidth}{!}{
\begin{tabular}{c c c c c c c c}
\hline \hline
\astro & Observation & Start Time & Stop time & Mode & Obs span & clean exposure$^a$\\
Instrument&ID& (yyyy-mm-dd hh:mm:ss) & (yyyy-mm-dd hh:mm:ss) & & (ks) & (ks)\\
\hline
LAXPC & 9000000634 & 2016-08-29 20:00:39 & 2016-08-30 20:47:37 & EA & 89.2 & 31.5 \\
SXT & 9000000634 & 2016-08-29 21:02:35 & 2016-08-30 19:57:57 & PC & 82.6 & 14.1 \\
\hline
\multicolumn{7}{l}{$^a$Net exposure of the persistent emission.}\\
\end{tabular}}
\label{obslog}
\end{table*}

\astro is India's first multi-wavelength (from optical to hard X-rays) astronomical mission. It was launched by Indian Space Research Organization in September 2015 \citep{Agrawal06, Singh2014}. It comprises of five scientific instruments -- Scanning Sky Monitor \citep[SSM: 2.5--10 keV;][]{Ramadevi18}, Soft X-ray Telescope \citep[SXT: 0.3--8.0 keV;][]{Singh17}, Large Area X-ray Proportional Counters \citep[LAXPC: 3--80 keV; ][]{Yadav2016, Agrawal17}, Cadmium Zinc Telluride Imager \citep[CZTI: 20--100 keV;][]{Rao17}, and Ultra-Violet Imaging Telescope \citep[UVIT: 1300--5500 $\angstrom$;][]{Tandon17}. For the current work, we have used data from SXT and LAXPC detectors only. 

\astro - LAXPC has capability for high time resolution $(10 \mu s)$ X-ray studies and covers a broad X-ray spectral band. It consists of three co-aligned proportional counters (LAXPC10, LAXPC20 and LAXPC30) which are sensitive to the X-ray photons in 3-80 keV energy range, with a total effective area of 6000 cm$^{-2}$ at 15 keV. It has a collimator with a $1^\circ \times 1^\circ$ field of view. For our analysis, data from LAXPC10 and LAXPC30 were not used due issues of high background and variable gain with the two instruments, respectively \citep{Agrawal17, Antia17}. LAXPC20 has $\sim$ 20\% spectral resolution at 30 keV \citep{Antia21}.

\astro - SXT is a focusing X-ray telescope fitted with a CCD in the focal plane. It can perform X-ray imaging and spectroscopy in the 0.3--8 keV energy range with an energy resolution of $\sim$150 eV. The peak effective area is $\sim$120 cm$^2$ in 0.8--2 keV range; $\sim$60 cm$^2$ in 2.5--5 keV and $\sim$7 cm$^2$ at 8 keV \citep{Singh2016, Singh17}. The on-axis FWHM of point spread function in the focal plane is $\sim 2$ arcmin.

The data from an \astro observation of \lmc (Observation Id G05\_115T01\_9000000634) made during 2016 August was analyzed for this work (See Table \ref{obslog} for observation details). During this observation, \lmc was observed for a span of $\sim$90 ks. The  Event Analysis (EA) mode data from LAXPC20 was processed by using the standard LAXPC software\footnote{\url{https://www.tifr.res.in/~astrosat\_laxpc/LaxpcSoft.html}} (\textsc{LaxpcSoft}: version 3.4.2). The light curves and spectra for the source and background were extracted from level 1 files by using the tool \textsc{laxpcl1}. The background in LAXPC is estimated from the blank sky observations \citep[see,][]{Antia17}. The photon arrival times in level 2 files were corrected to the solar system barycenter by using \textsc{as1bary}\footnote{\url{http://astrosat-ssc.iucaa.in/?q=data\_and\_analysis}} tool. 

For the spectral analysis, we have used the SXT data in addition to the LAXPC data. The SXT data of \lmc was taken in Photon Counting (PC) mode with time resolution of $\sim 2.37$ sec. Level 1 data was processed using \textsc{sxtpipeline} version 1.4b which generated the filtered level-2 cleaned event files. The cleaned event files from different orbits in SXT data were merged using \textsc{sxtevtmergertool}. A circular region of 12 arcmin radius was considered as a source region around the source location. Source events were not affected by pile-up as the observed average count rate ($\sim 2$ counts/s) was less than the threshold of 40 counts/s for pile-up in the PC mode. The spectra, image and light curve were extracted using the \textsc{xselect} v2.4m tool. The ancillary response file (ARF) was created with sxtARFModule using the arf file provided by SXT team. The response file (sxt\_pc\_mat\_g0to12.rmf) and the blank sky background spectrum file (SkyBkg\_comb\_EL3p5\_Cl\_Rd16p0\_v01.pha) were used which was provided by SXT team\footnote{\url{https://www.tifr.res.in/~astrosat_sxt/dataanalysis.html}}.

\section{ANALYSIS AND RESULTS}
\label{sec:result}

\begin{figure}
    \centering
  \includegraphics[width=\linewidth]{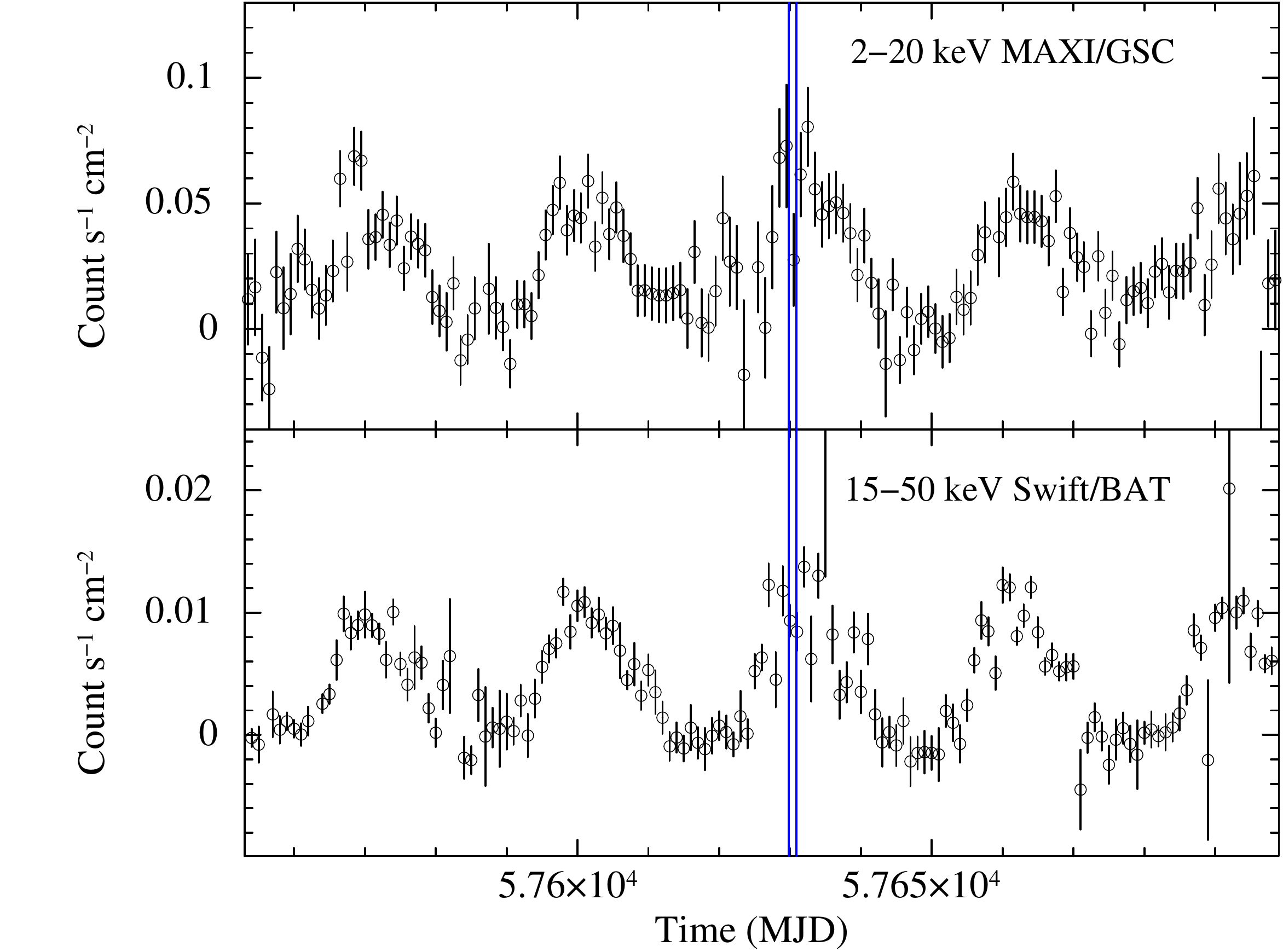}
\caption{Light curve of LMC X--4. Top Panel: The 2--20 keV \emph{MAXI}/GSC light curve. Bottom Panel: The 15--50 keV \emph{Swift}/BAT light curve. The span of \astro observation has been marked with blue vertical lines in both the panels.}
  \label{fig:lc}
\end{figure}

Figure \ref{fig:lc} shows the snapshot of long term light curve of LMC X--4. The top and bottom panels of this figure respectively correspond to 2--20 keV \emph{MAXI}/GSC \citep{Mihara2011} and 15--50 keV \emph{Swift}/BAT \citep{Krimm2013} light curves. The span of \astro observation is marked with vertical blue lines in both the panels. The observation was made during the high super-orbital phase of LMC X--4. 

\subsection{TIMING ANALYSIS}

\begin{figure}
  \centering
  \includegraphics[width=\linewidth]{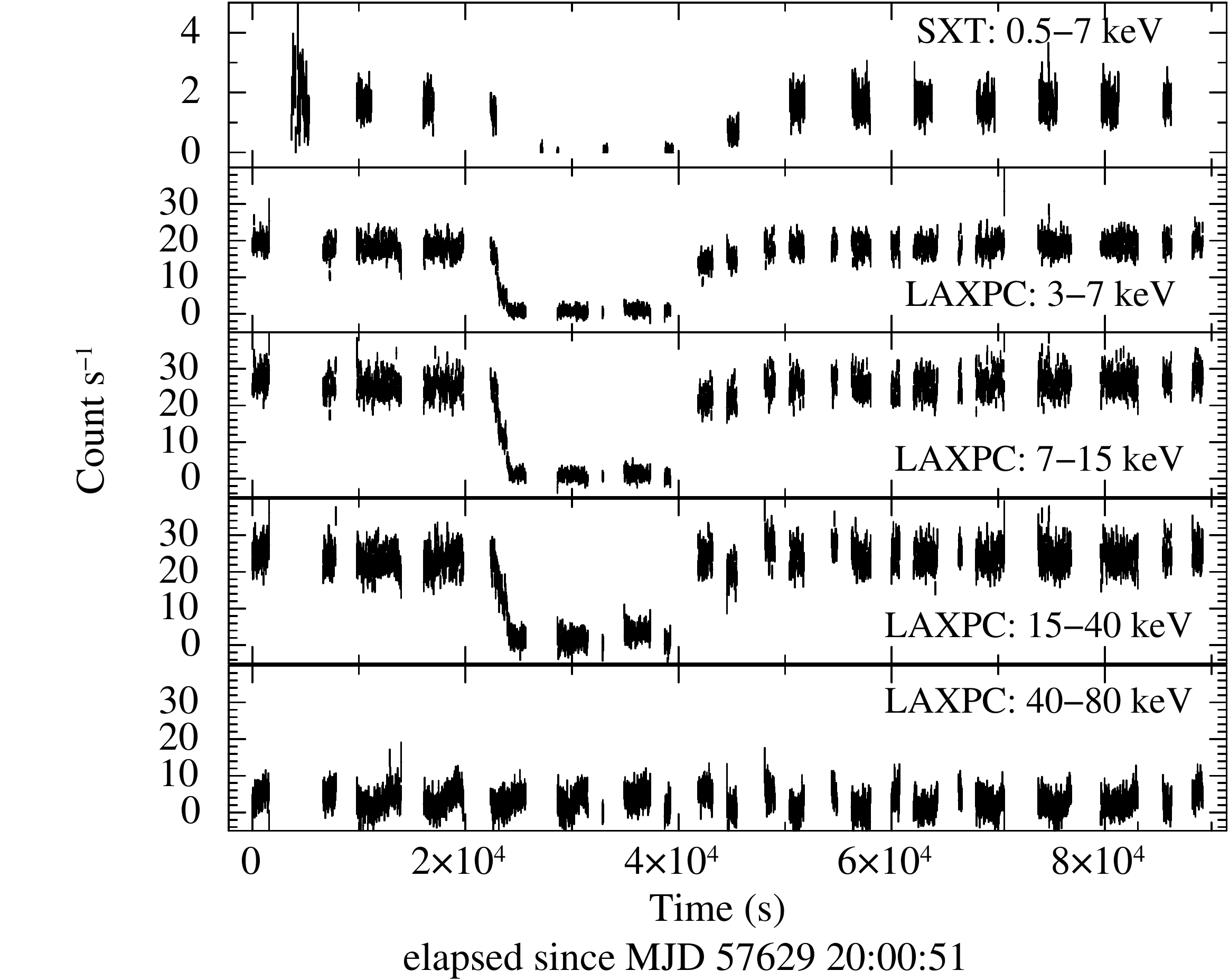}
  \caption{Background subtracted, barycenter corrected light curve of \lmc in different energy bands.}
  \label{fig:laxpc_lc}
\end{figure}

Figure \ref{fig:laxpc_lc} shows the background subtracted barycenter corrected light curve of \lmc obtained from SXT and LAXPC binned with 20 s. A complete X-ray eclipse lasting for about 18000 s is clearly visible. Starting from the top, the panels correspond to 0.5--7 keV, 3--7 keV, 7--15 keV, 15--40 keV and 40--80 keV, respectively. The count rate was very low in the 40--80 keV energy band and eclipse profile was not visible beyond 40 keV. The out of eclipse count rate remained constant and no flares were detected during the entire observation. For further analysis we have used only persistent count emission i.e. out of eclipse data.

\subsubsection{Power Density Spectrum}

We generated power density spectrum (PDS) from SXT and LAXPC light curves covering the intervals of persistent emission. The \textsc{powspec} tool of XRONOS sub-package of \textsc{ftools} \citep{Blackburn99} was used for this purpose. The SXT light curve binned at 2.37 sec was divided into stretches of 256 bins. 
The LAXPC light curve binned at 1 sec was divided into stretches of 1024 bins. The white noise was subtracted from all the PDS and they were normalized \citep{Leahy83} such that the integral between two frequencies corresponded to the squared r.m.s. fractional variability. The PDS from all the intervals were then averaged and re-binned geometrically in frequency. Owing to very low net exposure and a low count rate during the persistent emission, the PDS generated with SXT light curve had insufficient statistics and showed large error bars, therefore we have not used SXT data for timing analysis.

\begin{figure}
  \centering
 \includegraphics[width=0.9\linewidth]{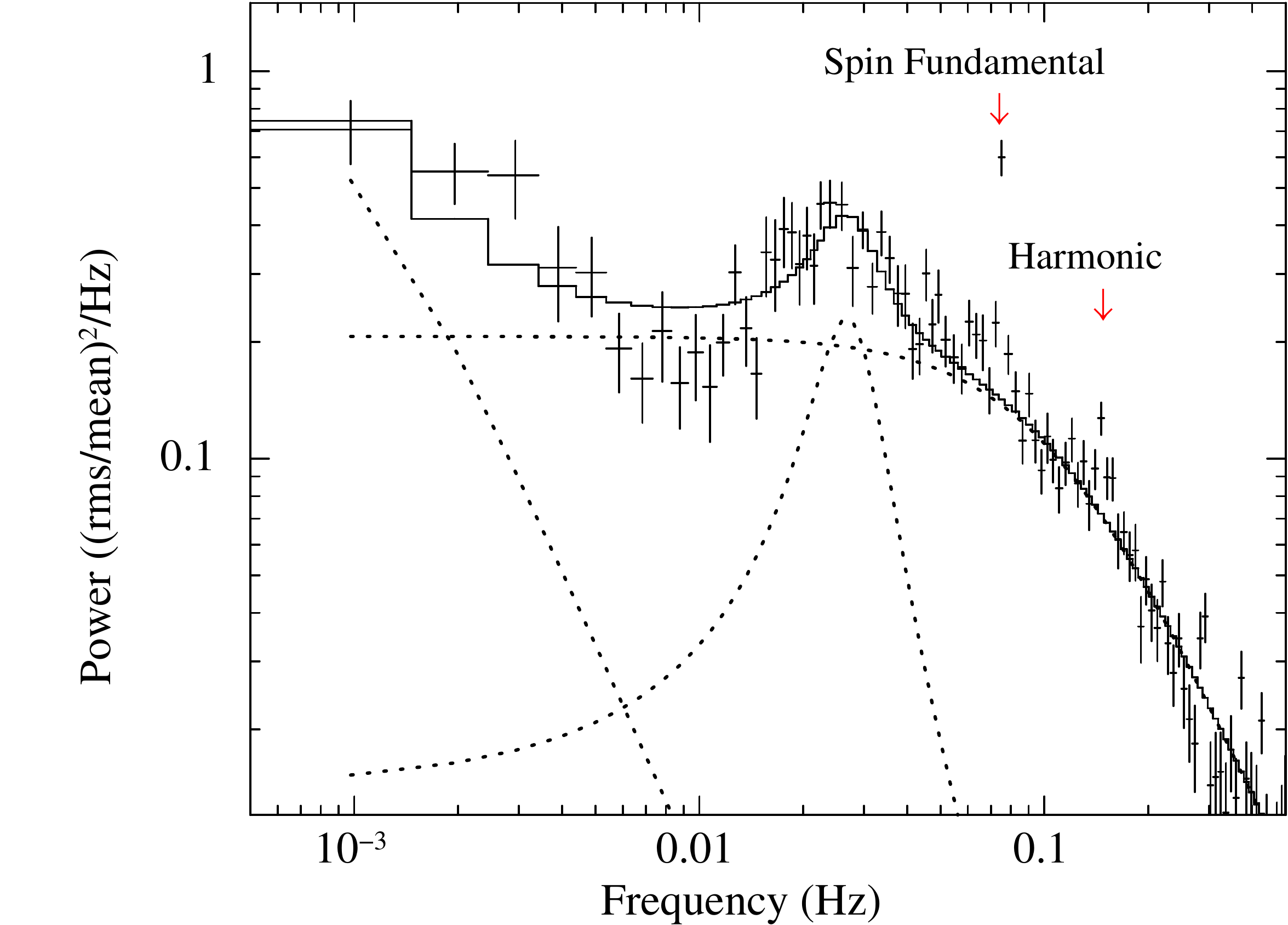}
  \caption{The 3--40 keV power density spectrum of LMC X--4. Thick solid line shows the best fit model. The dotted lines show individual components of the model.}
 \label{fig:pds}
\end{figure}

Figure \ref{fig:pds} shows the 3--40 keV PDS of \lmc\ obtained from LAXPC light curve. The PDS consisted of strong red noise, a QPO-like bump at 0.026 Hz, along with spin frequency peak at 0.074 Hz and its harmonic at 0.15 Hz (shown with red arrow marker in Figure \ref{fig:pds}). We modelled the PDS continuum with two Lorentzians centered at zero frequency. A third Lorentzian component was used to model the QPO feature at frequency of $\sim$26 mHz. 
The data points around the fundamental spin frequency and its harmonic were excluded from the fit. Table \ref{table:pds} gives the best fit model parameters (centre frequency of the Lorentzian $(\nu)$, full width half maximum (F.W.H.M. $(\delta \nu)$), quality factor $(Q = \frac{\nu}{\delta \nu})$ and r.m.s. amplitude of each Lorentzian components. The best fit model yielded a $\chi^2$ of 153 for 87 degrees of freedom (d.o.f). The error bars reported in this table are at $1 \sigma$ confidence level ($\Delta \chi^2$= 1.0). 

\begin{table}
\centering
\caption{Best fit parameters of 3--40 keV power density spectrum of \lmc\ during the persistent emission. The variation of parameters of the QPO feature as a function of energy is also mentioned. All errors reported in this table are at $1 \sigma$ confidence level ($\Delta \chi^2$=1.0).}
\label{table:pds}
\begin{tabular}{l l l l l}
\hline
Energy    & Frequency& F.W.H.M.& Quality & r.m.s. \\
 (keV)  &  $\nu$ (mHz)   &  $\delta \nu$ (mHz) & Factor &   (\%)\\
\hline
3--40 keV$^{a}$ & 0$^*$ & $1.7^{+0.9}_{-1.0}$ &   &  $3.9_{-0.4}^{+0.8}$ \\[1ex]
        &  0$^*$ & $212^{+10}_{-9}$  &   & $18.5 \pm 0.3 $ \\[1ex]
        & $26.8^{+1.3}_{-1.0}$ & $13.7^{+4.2}_{-3.2}$ & $1.96$ & $6.8_{-0.6}^{+0.7}$ \\[0.5ex]
\hline
\multicolumn{4}{l}{Variation of QPO parameters with energy}\\
\hline
3--6 & $26.6 \pm 1.5$    & $8.1 \pm 4.1$    & $3.3$ & $5.3 \pm 0.8$ \\
\hline
6--12 & $24.9 \pm 0.1$ & $10.4 \pm 4.4$    & $2.4$ & $5.95 \pm 0.79$ \\
\hline
12--20 & $25.7 \pm 1.7$ & 10$^{\rm fixed}$    & 2.57 & $5.9 \pm 0.6$ \\
\hline
20--40 & 24.1$\pm$1.3   & $10.4 \pm 3.8$    & $2.3$ & $7 \pm 1$ \\
\hline
\multicolumn{4}{l}{$^{a}$ $\chi^2$/dof = 153/87 for 3--40 keV PDS}\\
\multicolumn{4}{l}{$^{*}$This parameter was fixed at zero frequency}\\
\end{tabular}
\end{table}

\begin{figure}
  \centering
  \includegraphics[width=0.89\linewidth]{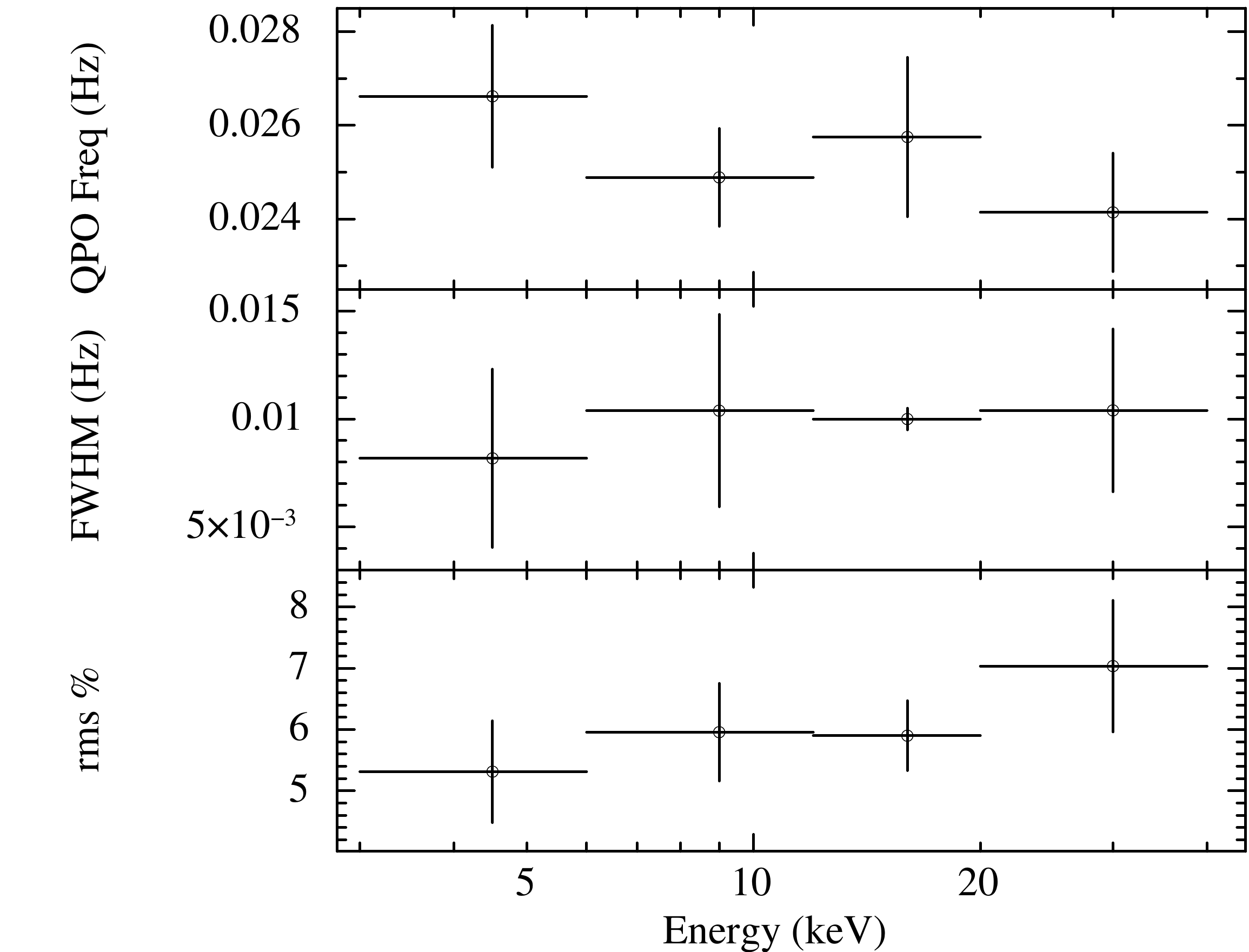}
  \caption{Variation of the QPO parameters with energy. For 12--20 keV energy range, the width (FWHM) of Lorentzian was fixed to 0.01 Hz.
  }
  \label{fig:qpo_variation}
\end{figure}

We also investigated the energy dependence of the 26 mHz QPO. The PDS was extracted in 3--6 keV, 6--12 keV, 12--20 keV, 20--40 keV and 40--80 keV energy bands. Above 40 keV, the QPO feature was insignificant. A summary of the QPO characteristics (centre frequency, F.W.H.M., quality factor and r.m.s. amplitude) as a function of energy is listed in Table \ref{table:pds} and shown in Figure \ref{fig:qpo_variation}. 
The quality factor does not show any clear trend with energy. Though the r.m.s. amplitude of the QPO appears to increase with energy, it is not statistically significant. 

\subsubsection{Pulsations}

The orbital period of \lmc is about 1.4 d and the pulses are expected to lose coherence within a short timescale of few thousands seconds. Therefore assuming a nearly circular orbit, the photon arrival time was corrected for the binary motion using ephemeris reported in \citet{Levine00}.%, Molkov2015}.

We used the epoch folding technique \citep{Leahy83} to determine the best period. The error in the spin period was estimated by using the bootstrap method described in \citet{Lutovinov12} and \citet{Boldin13}. Assuming that the count rate at any instant of time in the original light curve is $C_{orig}$ and it has an error of $Err_{orig}$ associated with it, we simulated 1000 light curves such that the new count rate ($C_{new}$) at each instant of time is given by Equation \ref{eqn3}. 
Here, $\alpha$ is a uniformly distributed variate between [-1,1].

\begin{equation} \label{eqn3}
C_{new} = C_{orig} + \alpha Err_{orig}
\end{equation}

Spin period was obtained for each of the simulated light curves by using the epoch folding technique. We obtained a standard deviation of $1.6 \times 10^{-5}$ sec in the best spin period distribution. This number was taken as error in pulse period. As a result, the best pulse periods of \lmc for \astro observation obtained is 13.501606 (16) sec for epoch 57630.0 MJD. 

\begin{figure}
  \centering
  \includegraphics[width=\linewidth]{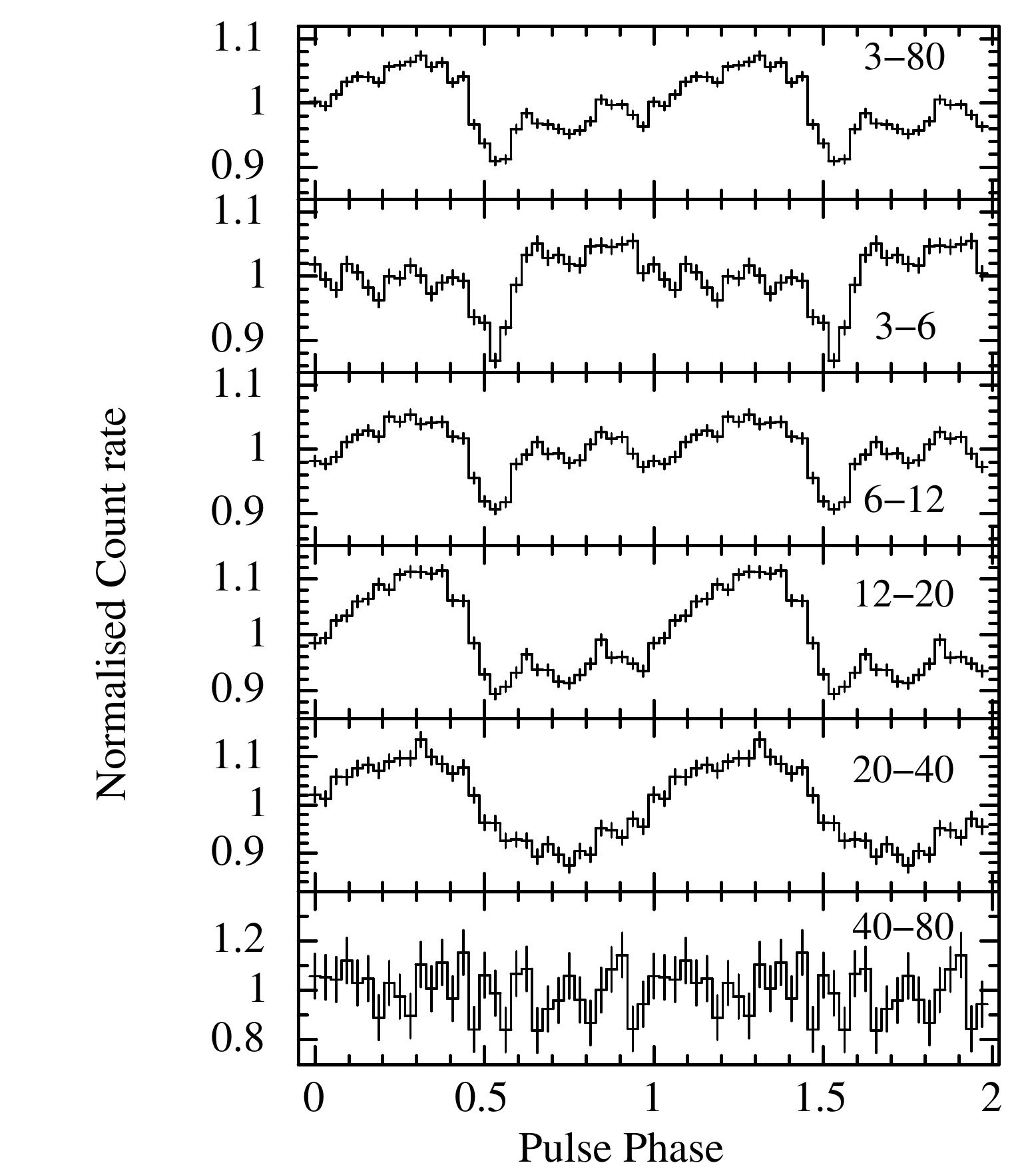}
  \caption{Energy resolved pulse profile of \lmc in energy bands 3--80, 3--6, 6--12, 12--20, 20--40 and 40--80 keV, respectively.}
  \label{fig:pulseprofile}
\end{figure}

We created energy resolved pulse profiles using the folding technique with the best period obtained for this observation. Figure ~\ref{fig:pulseprofile} shows the energy resolved pulse profiles of \lmc in the energy range of 3--80, 3--6, 6--12, 12--20, 20--40 and 40--80 keV, respectively. The observed pulse profiles are similar to that reported with data from \emph{Ginga} \citep{Woo96}, \emph{RXTE} \citep{Levine00}, \emph{Beppo}SAX \citep{Naik04}, \emph{Suzaku} \citep{Hung10} and \emph{NuSTAR} \citep{Shtykovsky17}. The low energy pulse profiles (3--6 keV and 6--12 keV) comprise of complex structures showing narrow dip around 0.5--0.6 pulse phase. These dips could be due to obscuration by accretion stream \citep{Paul02, Naik04, Beri2017}. Dips are not observed above 20 keV and the profile is a smooth sinusoidal in the energy range 20--40 keV. Due to very low signal-to-noise ratio (See Fig. \ref{fig:bkg}), we did not detect pulsed emission in 40--80 keV energy range. 

\begin{figure}
  \centering
 \includegraphics[width=0.9\linewidth]{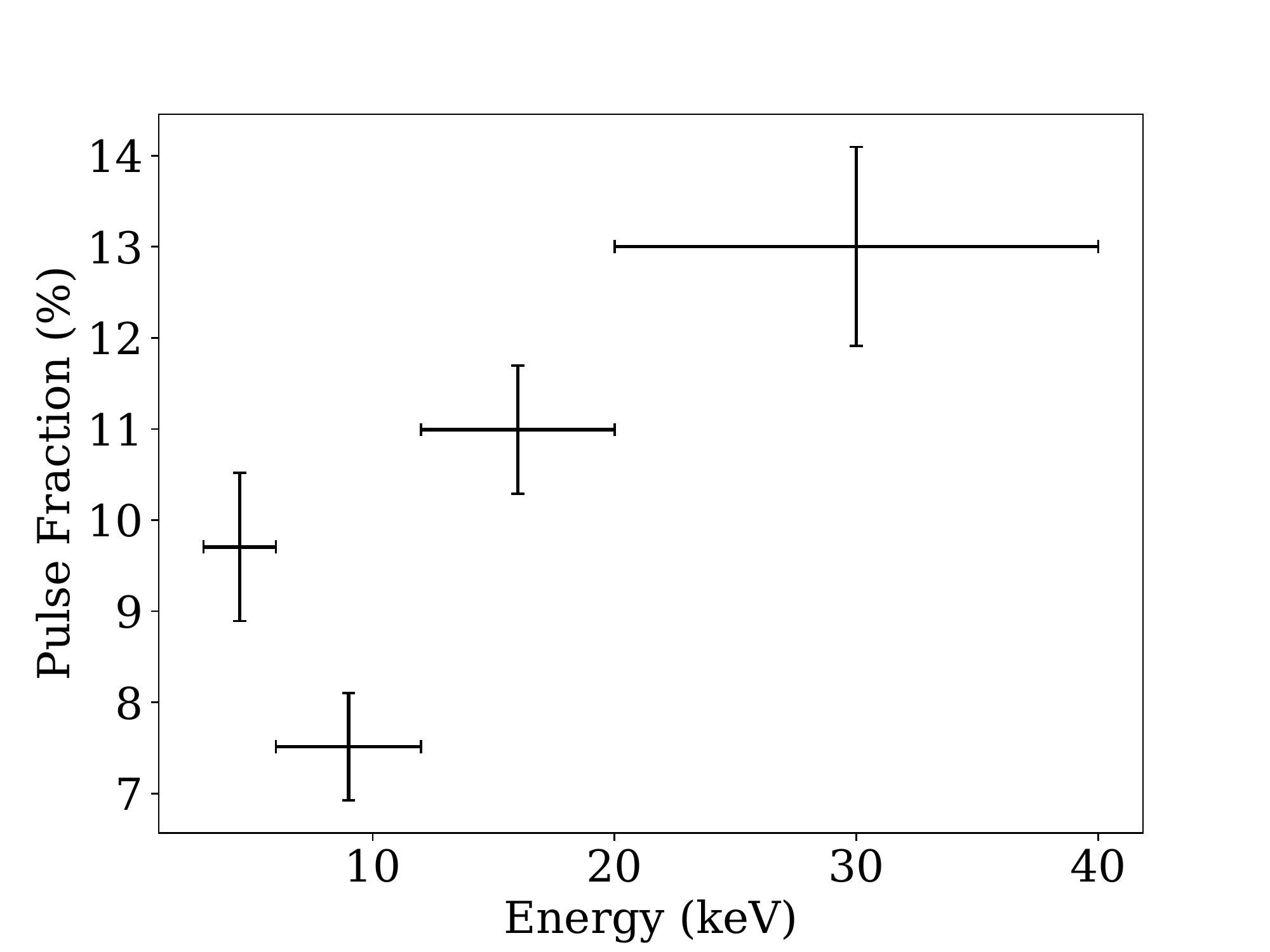}
  \caption{Energy dependence of the pulsed fraction of LMC X--4.}
  \label{fig:pulsefraction}
\end{figure}

Figure \ref{fig:pulsefraction} presents the energy dependence of the pulsed fraction $P_{\rm F} \left( = \frac{I_{\rm max} - I_{\rm min}}{I_{\rm max} + I_{\rm min}}\right)$, where $I_{\rm max}$ and $I_{\rm min}$ are the maximum and minimum intensities of the pulse profile in the corresponding energy range, respectively. 
From $\sim 10\%$ in the 3--6 keV energy range, the pulse fraction decreased to $\sim 7\%$ in 6--12 keV energy range. Thereafter, the pulse fraction shows a marginally increasing trend with increasing energy, $\sim 11\%$ and $\sim 13\%$ in 12--20 keV and 20--40 keV, respectively. We obtained a $\chi^2$ of 10.9 for 2 dof for a linear trend compared to a $\chi^2$ of 26 for 3 dof for a constant model. The 3-6 keV pulse profile shows narrow dip (due to obscuration from accretion stream) which causes increase in pulse fraction. Above 6 keV, pulse fraction show an increase with energy. Similar decrease/dip in pulse fraction was also observed with one of the \emph{NuSTAR} observation of \lmc \citep{Shtykovsky18}. The increase in pulse fraction with energy has been observed in several bright X-ray pulsars \citep{Lutovinov09, Shtykovsky17}.

\subsection{Spectral Analysis}

To study the broadband emission spectra of LMC X--4, we performed combined spectral analysis of SXT and LAXPC20. Spectral fitting was performed by using the \textsc{xspec} v 12.12.0 of the Heasoft 6.29 package \citep{Arnaud96}. A systematic uncertainty of 1\% was used during spectral fitting. A background comparison with the observed signal from SXT and LAXPC is shown in fig. \ref{fig:bkg}. The LAXPC20 spectrum above 25 keV was ignored because of larger uncertainty in background estimation around the K-fluorescence energy of Xe at 30 keV \citep{Antia17, Sharma2020, Beri2021}, which makes it difficult to measure spectrum of faint sources with LAXPC at higher energies. The SXT and LAXPC spectrum were grouped using \textsc{grppha} to have a minimum count of 20 counts per bin. We added a constant component to represent the cross-calibration constant between the SXT and the LAXPC20 instruments. We also applied a gain correction for the SXT. The gain slope was fixed to 1.0 and the gain offset was allowed to vary. We observed a gain offset of $\sim 32$ eV.

We fitted the 0.5--25 keV SXT+LAXPC spectrum by power law with high energy cutoff ($\texttt{highecut} \times \texttt{powerlaw}$ in \textsc{xspec}) with line of sight absorption (\texttt{tbabs} in \textsc{xspec}). \lmc does not show significant intrinsic absorption. The absorption column density in a direction towards \lmc is low. It is close to the value of galactic absorption. So initially we fixed it to $8 \times 10^{20}$ cm$^{-2}$ \citep{HI4PI}. Later we allowed the $N_H$ parameter to vary and found the upper limits of $N_H < 5 \times 10^{20}$ atoms cm$^{-2}$ \citep{Paul02}.

\begin{figure}
  \centering
  \includegraphics[width=\linewidth]{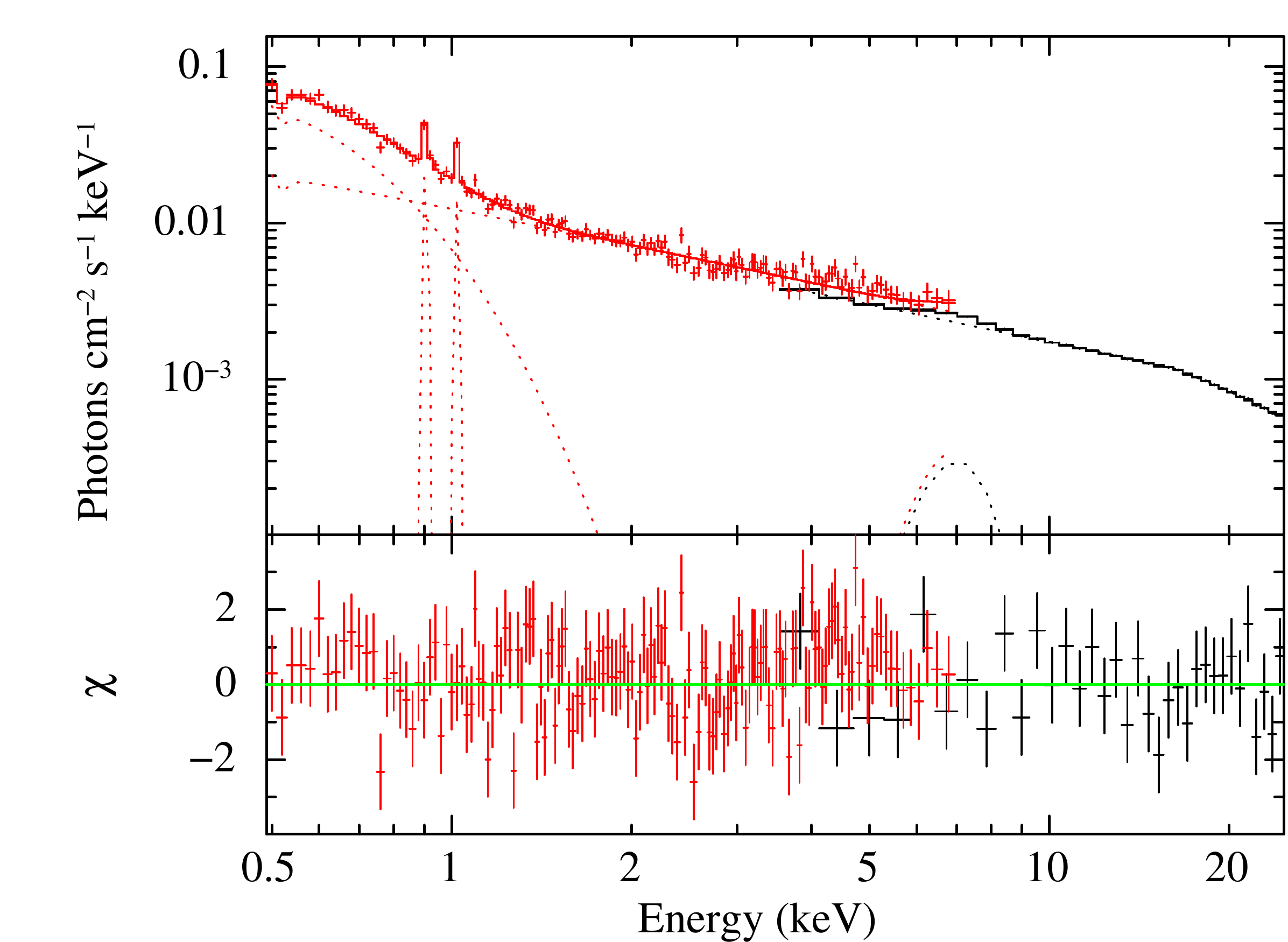}
  \caption{The best fit 0.5--25 keV SXT+LAXPC spectrum of LMC X--4. The bottom panel shows the residuals ($\chi$=(data--model)/error) with respect to best fit model. The figure has been rebinned for plotting purpose.}
  \label{fig:spectra}
\end{figure}

The fit obtained with \texttt{highecut $\times$ powerlaw} model was not good. The residuals showed significant soft excess at low energies and emission feature at 6--7 keV. To model the emission feature, we added a Gaussian component to the model. 
The soft excess was modelled with addition of a soft thermal component (\texttt{bbodyrad}). Addition of emission line and soft thermal component significantly improved the fit. The residuals around 1 keV from SXT indicated emission features. Therefore, two Gaussian emission line components were added at 0.9 keV and 1.02 keV for Ne \textsc{ix} and Ne \textsc{x} respectively \citep{Neilsen09, Hung10}. The best fit 0.5--25 keV persistent spectrum of \lmc is shown in Figure \ref{fig:spectra} and the best fit parameters are listed in Table \ref{table:spectra}.

\begin{table}
\centering
\caption{Best fit spectral parameters of \lmc during the persistent emission. All errors and upper limits reported in this table are at $90\%$ confidence level ($\Delta \chi^2$=2.7).}
\label{table:spectra}
\resizebox{0.9\linewidth}{!}{
\begin{tabular}{l l l}
\hline
Model & Parameters   & SXT+LAXPC\\
\hline
TBabs & $N_H$ ($10^{22}$ cm$^{-2}$) & $<0.05$\\
\\
BBodyrad & $kT_{\rm BB}$ (keV) & $0.14 \pm 0.01$ \\[1ex]
        & norm$_{\rm BB}$ & $7700^{+4600}_{-2700}$ \\[1ex]
        & $R_{\rm BB}$ (km)$^c$ & $440^{+130}_{-80}$ \\
\\
HighEcut & $E_{\rm cutoff}$ (keV) & $16.14^{+0.43}_{-0.34}$\\[1ex]
       & $E_{\rm fold}$ (keV) & $22.0^{+1.4}_{-1.0} $\\
\\      
Powerlaw & $\Gamma$  & $0.799 \pm 0.012 $ \\[1ex]
        & Norm & $0.0109 \pm 0.0003$\\
\\
Gaussian (Fe K) & $E$ (keV)  & $7.00 \pm 0.16$ \\[1ex]
        & $\sigma$ (keV) & $0.85 \pm 0.23$ \\[1ex]
        & Eqw (eV) & $291^{+60}_{-66}$ \\[1ex]
        & Norm ($10^{-4}$) & $6.6^{+1.4}_{-1.1}$ \\
\\
Gaussian (Ne \textsc{ix} He$_\alpha$) & $E$ (keV) & $0.904^{+0.018}_{-0.021}$ \\[1ex]
        & $\sigma$ (keV) & $0.005^{fixed}$ \\[1ex]
        & Eqw (eV) & $19$ \\[1ex]
        & Norm ($10^{-4}$)  & $4.07^{+1.5}_{-1.6}$ \\
\\
Gaussian (Ne \textsc{x} L$y_\alpha$) & $E$ (keV) & $1.02 \pm 0.02$ \\[1ex]
        & $\sigma$ (keV) & $0.005^{fixed}$ \\[1ex]
        & Eqw (eV) & $17.2$ \\[1ex]
        & Norm ($10^{-4}$) & $2.9 \pm 1.1$ \\
\\
Constant & $C_{\rm LAXPC}$ & $1^{fixed}$ \\
        & $C_{\rm SXT}$ & $1.155^{+0.029}_{-0.023}$ \\
\\
Unabs. Flux$^{a}$ & $F_{0.5-25 \rm ~keV}^{\rm total}$ & $6.6 \times 10^{-10}$\\[1ex]
 & $F_{0.1-100 \rm ~keV}^{\rm total}$ & $1.208 \times 10^{-9}$\\[1ex]
 & $F_{0.1-100 \rm ~keV}^{\rm pow}$ & $1.17 \times 10^{-9}$\\[1ex]
 & $F_{0.1-100 \rm ~keV}^{\rm BB}$ & $2.98 \times 10^{-11}$\\[1ex]

X-ray Luminosity$^{b}$ & $L_{0.5-25 \rm ~keV}$ & $1.97 \times 10^{38}$ \\[1ex]
& $L_{0.1-100 \rm ~keV}$ & $3.6 \times 10^{38}$ \\[1ex]

\\            
          & $\chi^2/{\rm dof}$ & 466.8/412 \\

\hline
\multicolumn{3}{l}{$^{a}$Unabsorbed Flux in the units of \erg.}\\
\multicolumn{3}{l}{$^{b}$X-ray luminosity in the units of \lum.}\\
\multicolumn{3}{l}{$^c R_{\rm BB}$ is obtained assuming no color correction.} \\
\end{tabular}}
\end{table}

We have determined an average unabsorbed 0.5--25 keV flux of $6.6 \times 10^{-10}$ \erg. This corresponds to an unabsorbed X-ray luminosity of $L=1.97 \times 10^{38}$ \lum\ for a distance of 50 kpc \citep{Piet13}.  Assuming absence of other spectral features in a wider energy range, the extrapolated 0.1--100 keV flux was calculated to be $1.2 \times 10^{-9}$ \erg, which translates to an X-ray luminosity of $3.6 \times 10^{38}$ \lum.

\section{DISCUSSION}
\label{sec:discuss}

In this work, we have presented the results of broadband timing and spectral analysis of HMXB \lmc by using \astro data of 2016 August. The source was in its high state of super-orbital motion during this observation. We conclude the following.
\begin{enumerate}
  \item Detection of energy dependent 26 mHz QPO in a wide energy band of 3--40 keV, 
  \item Pulse period at 13.501606 (16) sec and energy dependent pulse profiles,
  \item 0.5--25 keV energy spectrum comprised of absorbed high energy cutoff power law ($\Gamma \sim 0.8$) with high energy cutoff at $\sim 16$ keV, a soft thermal component at $0.14$ keV and Gaussian components corresponding to Fe K$_\alpha$, Ne \textsc{ix} and Ne \textsc{x} spectral lines.
  
\end{enumerate}

QPOs have been detected in the power spectrum of several accreting high magnetic field X-ray pulsars \citep{Angelini89, Takeshima91, Finger96, Paul98, Kaur07} and Low Mass X-ray Binaries \citep[LMXBs;][]{Zhang96, Makishima99, Kaur08, Jain10}. The QPO characteristics, especially their frequency and their evolution with time and energy, have been used to understand the physical processes that lead to QPO generation, the size of the emitting region and their connection with the spectral parameters \citep{Finger98, Bult15}.  

In LMC X--4, the QPO r.m.s. is seen to increase with energy with a slope of 0.06(5) \% per keV in a broad energy band of 3--40 keV. This indicates that these aperiodic oscillations are not driven by the cold disc component. It is most likely due to variation in the hot plasma. This is also corroborated by the fact that the source luminosity is dominated by the power-law emission (integrated flux of $1.17 \times 10^{-9}$ \erg) rather than the black-body component (integrated flux of $2.98 \times 10^{-11}$ \erg).
Although \citet{Rikame22} have reported a decrease in r.m.s. variability with an increase in energy; the r.m.s. variability in the low energy band (0.3--1 keV; thermal dominated) was 10.1\% and in the high energy band (1--10 keV; non-thermal dominated) was 5.3\%. The difference in r.m.s. variability in the 0.3--1 keV band compared to the rest of the energy bands could be due to difference in the origin of X-ray radiation at these energies. The soft X-ray emission (below 1 keV) is dominated by the reprocessed emission from the structure around the neutron star (such as accretion disc) whereas the hard X-ray emission is coming from the accretion column above the neutron star poles. The r.m.s. variability of the QPO in this work as well as in \citet{Rikame22} for energies above 1 keV is similar.

\lmc has been extensively investigated by several X-ray missions. But in order to ascertain the presence of QPO feature, we analysed the light curves from these missions without going into a detailed analysis. \emph{XMM}-Newton observed \lmc on six occasions. But the $\sim 27$ mHz QPO has been detected in only two observations \citep[see,][for more details]{Rikame22}. \emph{Beppo}SAX-MECS has observed \lmc twice but QPO feature was not detected in either of them. \lmc was observed by \emph{Suzaku} thrice. A QPO feature at $\sim 31$ mHz with r.m.s. of 4.5\% was detected in only one observations (obs-ID = 702037010) and that too with $3\sigma$ significance. There was no detection of a QPO in the longest observation of \emph{RXTE}. Non-detection of QPOs in most observations clearly suggests that presence of a QPO in \lmc is a transient phenomena, similar to several other accreting pulsars such as Cen X-3, KS 1947+300 and 1A 1118-615 \citep{Raichur2008, James2010, Nespoli2011}.

The pulse profiles of LMC X-4 are significantly energy dependent. The pulse profiles in the soft X-ray energies (3--6 keV and 6--12 keV) showed complex dip-like features, possibly due to absorption from the accretion stream \citep{Paul02, Naik04, Beri2017}. Similar features are also reported in the pulse profile of other pulsars \citep[e.g. see,][]{Naik2008, Naik2013}. At higher energies, profiles are single-peaked, smooth and show an almost sinusoidal behaviour. 

The energy spectrum of \lmc is generally described by model comprising of a power-law with high-energy cutoff \citep{Woo96, Barbera01, Hung10}. The Comptonization model has also been used by some authors \citep{Barbera01}. We have modelled the 0.5--25 keV energy spectrum with a power law continuum ($\Gamma \sim 0.8$) and a high energy cutoff of $\sim$ 16 keV. Similar numbers have been reported with previous missions (\emph{Ginga} and \emph{ROSAT} \citep{Woo96}, \emph{Beppo}SAX \citep{Barbera01}, and \emph{NuSTAR} \citep{Shtykovsky17}. 

From the spectral fits, we have found that the power-law emission dominates the black-body component above 1 keV. Following \citet{Paul02} and \citet{Hickox04}, it is possible that the power-law emission from the neutron star is reprocessed by the inner accretion disc from where the black-body radiation is emitted. Using the normalization of \textrm{bbodyrad}, and assuming a circular emission geometry, the radius of the black-body emission region is estimated to be $440^{+130}_{-80}$ km for distance of 50 kpc. 

From the SXT spectrum, we observed a broad emission line from Fe K$_\alpha$ and narrow emission lines Ne \textsc{ix} and Ne \textsc{x} with equivalent width of 291 eV, 19 eV and 17 eV, respectively. From the \emph{NuSTAR} observations, \citet{Shtykovsky17} determined an equivalent width of the iron line to be $\sim 158$ eV. \citet{Neilsen09} observed many narrow H-like and He-like emission lines from elements O, N, Ne and Fe. They proposed that Ne \textsc{ix} originated from the stellar winds while Ne \textsc{x} had its origin in the outer accretion disc. These lines were also observed with \emph{Suzaku} spectra \citep{Hung10} and with \emph{XMM}-Newton+\emph{NuSTAR} spectra \citep{Brumback2020}. The equivalent width of the observed Fe emission line is consistent with that reported by \citet{Naik03} and \citet{Neilsen09} during the high state of LMC X--4.

Using $M = 1.25 M_\odot$ \citep{Meer07} for the neutron star and $R_{\rm BB} \sim 440$ km, we find the velocity at the inner accretion disc to be $\sim 2 \times 10^4$ km s$^{-1}$, which corresponds to Doppler shift of $\sim 0.4$ keV for Fe K$_\alpha$. The line width observed from spectral fit was $\sim 0.8$ keV. This broad Fe line could be Doppler broadened Fe K$_\alpha$ emitted from the inner accretion disc due to the high velocity or produced by emission from Fe in a range of ionization states \citep{Neilsen09, Hung10}. 

\section*{Acknowledgments}
This work has made use of data from the \astro mission of the Indian Space Research Organisation (ISRO), archived at the Indian Space Science Data Centre (ISSDC). We  thank the LAXPC Payload Operation Center (POC) and the SXT POC at TIFR, Mumbai for providing necessary software tools. We have also made use of data and software provided by the High Energy Astrophysics Science Archive Research Center (HEASARC), which is a service of the Astrophysics Science Division at NASA/GSFC. We also thank the anonymous referee for insightful comments and suggestions.

%%%%%%%%%%%%%%%%%%%%%%%%%%%%%%%%%%%%%%%%%%%%%%%%%%

\section*{Data Availability}

Data used in this work can be accessed through the Indian Space Science Data Center (ISSDC) at 
\url{https://astrobrowse.issdc.gov.in/astro\_archive/archive/Home.jsp}.



\bibliographystyle{mnras}
\bibliography{refs}

%\begin{thebibliography}{99}
%\end{thebibliography}

%%%%%%%%%%%%%%%%%%%%%%%%%%%%%%%%%%%%%%%%%%%%%%%%%%

%%%%%%%%%%%%%%%%% APPENDICES %%%%%%%%%%%%%%%%%%%%%

\appendix

\section{Source and Background Comparison}

\begin{figure}
    \centering
  \includegraphics[width=\linewidth]{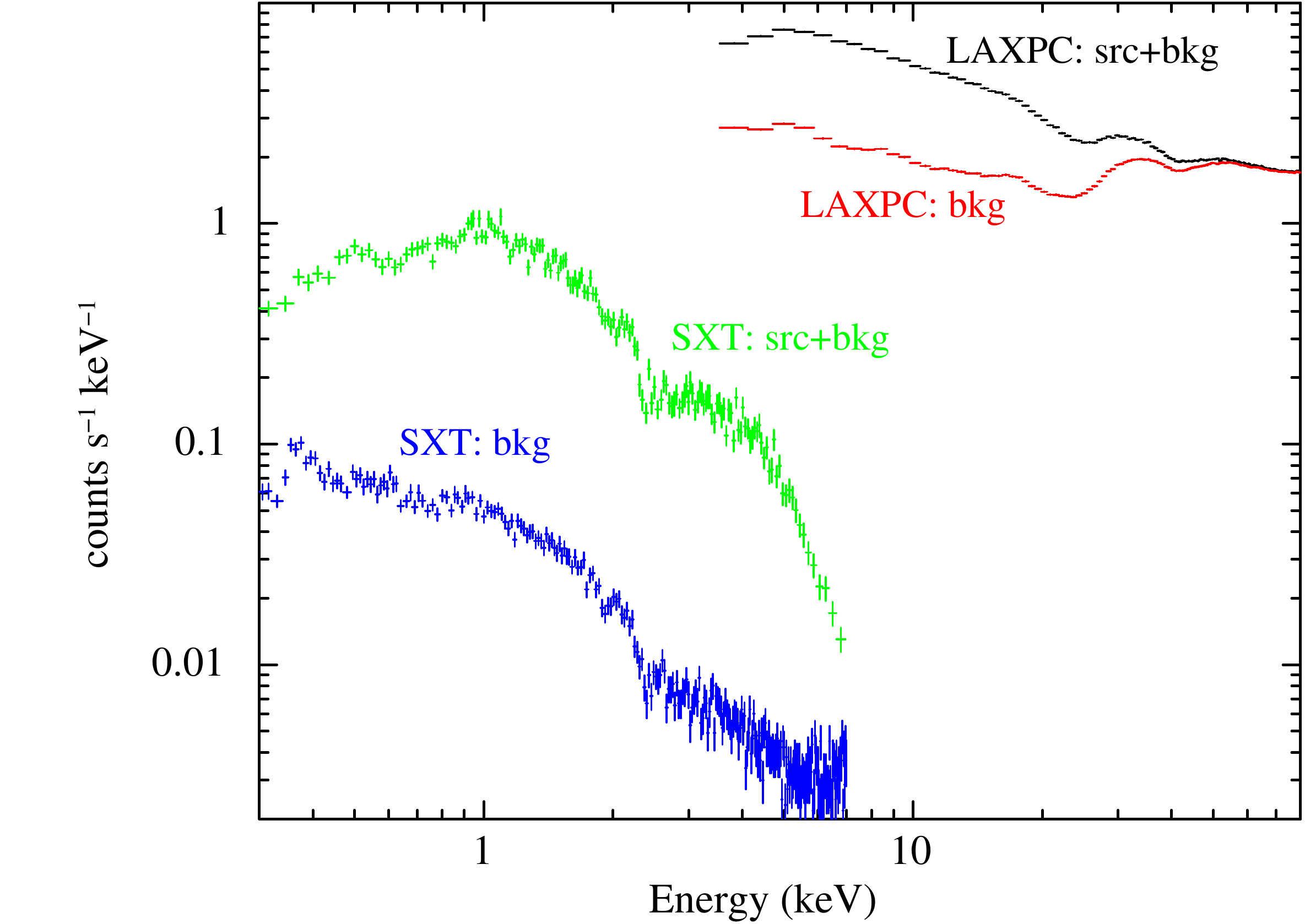}
\caption{The comparison of source+background and background only spectrum from SXT and LAXPC for the persistent region. SXT spectra has been rebinned for plotting purpose. Black and green correspond to source+background spectra for LAXPC and SXT, respectively, and the red and blue correspond to background spectra for LAXPC and SXT, respectively. Although the signal to noise ratio indicates that the energy range can be extended to higher energies, but systematic features dominate the spectrum \citep[see e.g.,][]{Antia17, Beri2021}.}
  \label{fig:bkg}
\end{figure}

%%%%%%%%%%%%%%%%%%%%%%%%%%%%%%%%%%%%%%%%%%%%%%%%

\bsp	% typesetting comment
\label{lastpage}
\end{document}